\documentclass[conference,a4paper]{APSIPA2017}
\usepackage{multirow}
\usepackage{graphicx}
\usepackage{amsmath}
\usepackage[psamsfonts]{amssymb}
\usepackage{amsxtra}
\usepackage{threeparttable}
\usepackage{amsmath,amssymb}

\usepackage{txfonts}
\usepackage{bm}
\usepackage{color}
\usepackage{CJK}

\begin{document}

\title{Deep Speaker Verification: Do We Need End to End?}

\author{%
\authorblockN{%
Dong Wang\authorrefmark{2}, Lantian Li\authorrefmark{2}, Zhiyuan Tang\authorrefmark{2}, Thomas Fang Zheng\authorrefmark{2}
}

\authorblockA{%
\authorrefmark{2}
Center for Speech and Language Technologies, Research Institute of Information Technology \\
Department of Computer Science and Technology, Tsinghua University, China
}


%
}

\maketitle
\thispagestyle{empty}

\makeatletter{\renewcommand*{\@makefnmark}{}
\footnotetext{D.W. and L.L. are joint first authors with equal contribution. }\makeatother}

\begin{abstract}

End-to-end learning treats the entire system as a whole adaptable black box, which, if sufficient data are available, may
learn a system that works very well for the target task. This principle has recently been applied to several prototype research on speaker
verification (SV), where the feature learning and classifier are learned together with an objective function that is consistent
with the evaluation metric.
An opposite approach to end-to-end is feature learning, which firstly trains a feature learning model, and then constructs a back-end classifier separately to perform SV. 
Recently, both approaches achieved significant performance
gains on SV, mainly attributed to the smart utilization of deep neural networks. However, the two approaches have not been
carefully compared, and their respective advantages have not been well discussed.
In this paper, we compare the end-to-end and feature learning approaches on a text-independent SV task. Our experiments on a
dataset sampled from the Fisher database and involving 5,000 speakers demonstrated that the feature learning approach outperformed the end-to-end approach.
This is a strong support for the feature learning approach, at least with data and computation resources similar to ours.

\end{abstract}

\section{Introduction}
\label{sec:intro}

Speaker verification (SV) is an important biometric recognition technology and has gained great popularity in a wide range of applications,
such as access control, transaction authentication, forensics and personalization.
After decades of research, SV has gained significant performance improvement, and has been deployed in some practical applications~\cite{campbell1997speaker,reynolds2002overview,kinnunen2010overview,hansen2015speaker}.
However, SV is still a very challenging task, mainly attributed to the large uncertainty caused by the complex convolution
of various speech factors, especially phone content and channel~\cite{zheng2014overview}.

Most of the existing successful SV approaches rely on probabilistic models to factorize speech signals into
factors related to speakers and other variations, especially the phone content.
A classical probabilistic model is the Gaussian mixture model-universal background model (GMM-UBM)~\cite{Reynolds00}, where
the speaker factor is assumed to be an additive component to the phone variation (represented by Gaussian components). This model was extended to a low-rank formulation, leading to
the joint factor analysis (JFA) model~\cite{Kenny07} and its `simplified' version, the famous i-vector model~\cite{dehak2011front}.
To further improve speaker-related discrimination,  various discriminative back-end models have been proposed, e.g.,
metric learning~\cite{schultz2004learning}, linear discriminant analysis (LDA)~\cite{dehak2011front} and its probabilistic version, PLDA~\cite{Ioffe06}. A DNN-based i-vector model was also proposed~\cite{Kenny14,lei2014novel}, where a phonetic deep neural
network (DNN) is used to enhance the factorization for speaker factors by providing phonetic information.

Recently, the deep learning approach has gained much attention in the SV
research. Different from the probabilistic methods, these deep SV methods utilize various DNN structures to learn
speaker features. This can be regarded as a neural-based speech factorization, which is deep, non-linear
and non-Gaussian.
The initial work towards the deep SV was proposed by Ehsan and colleagues~\cite{ehsan14}. They constructed a DNN
model with $496$ speakers in the training set as the targets. The frame-level features were read from the activations of the
last hidden layer, and the utterance-level representations (called `d-vector') were obtained by averaging over the frame-level
features. In evaluation, the decision score was computed as a simple cosine distance between the d-vectors of the enrollment
utterance and the test utterance.
This preliminary results triggered much interest on deep SV. Many researchers quickly noticed the inferior performance
of this approach compared to the counterpart i-vector model might be caused by the naive back-end model, i.e., the
frame averaging and the cosine-based scoring. An `end-to-end approach' was developed that learns the back-end model
together with the feature learning~\cite{heigold2016end,zhang2017end,snyderdeep16,li2017}.

Another approach focuses on learning speaker features, leaving the back-end model as a separate component.
The idea is that if the feature learning is strong enough, the back-end model issue will be naturally solved.
Our group followed this direction, and found that a simple CT-DNN structure can learn speaker features very well~\cite{li2017deep}.

These two deep SV approach: end-to-end and feature learning, however, have not been carefully compared.
In this paper, we present a comparative experimental study for the two deep SV
approaches. Based on a training database consisting of $5,000$ speakers, we found
the feature learning approach performs consistently better than the end-to-end
approach. This result is a strong support for the
feature learning approach, at least in conditions similar to our experiment.

The rest of this paper is organized as follows. Section~\ref{sec:model} presents
the two deep SV learning approaches in detail.
The comparative experiments are presented in Section~\ref{sec:exp}, and Section~\ref{sec:conl}
concludes the paper.

\section{Deep speaker verification models}
\label{sec:model}

This section presents the model structures of the feature learning approach and the end-to-end approach used in
our study. The former was proposed by our group~\cite{li2017deep}, and the
latter was proposed by Snyder et al.~\cite{snyderdeep16}.

\subsection{Feature learning model}

\begin{figure*}[htp]
\centering
\includegraphics[width=0.95\linewidth]{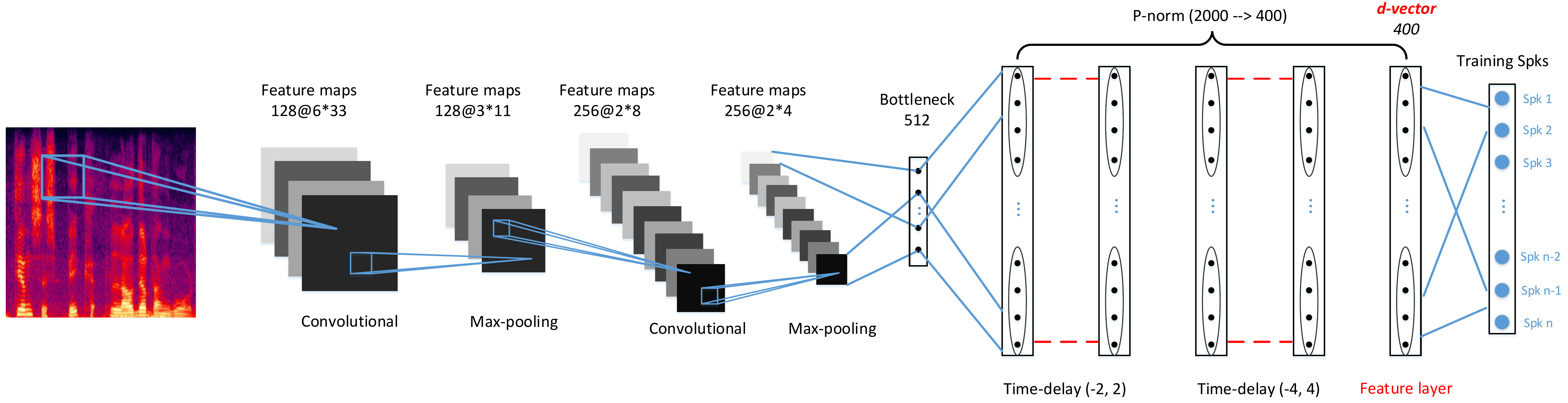}
\caption{The DNN structure of the deep feature learning system~\cite{li2017deep}.}
\label{fig:ctdnn}
\end{figure*}

The DNN structure of the feature learning system is illustrated in Fig.~\ref{fig:ctdnn}.
It consists of a convolutional (CN) component and a time-delay (TD) component, connected
by a bottleneck layer. The frame-level
speech features are read from the last hidden layer (feature layer).
More details can be found in~\cite{li2017deep}.

\begin{figure*}[htp]
\centering
\includegraphics[width=0.95\linewidth]{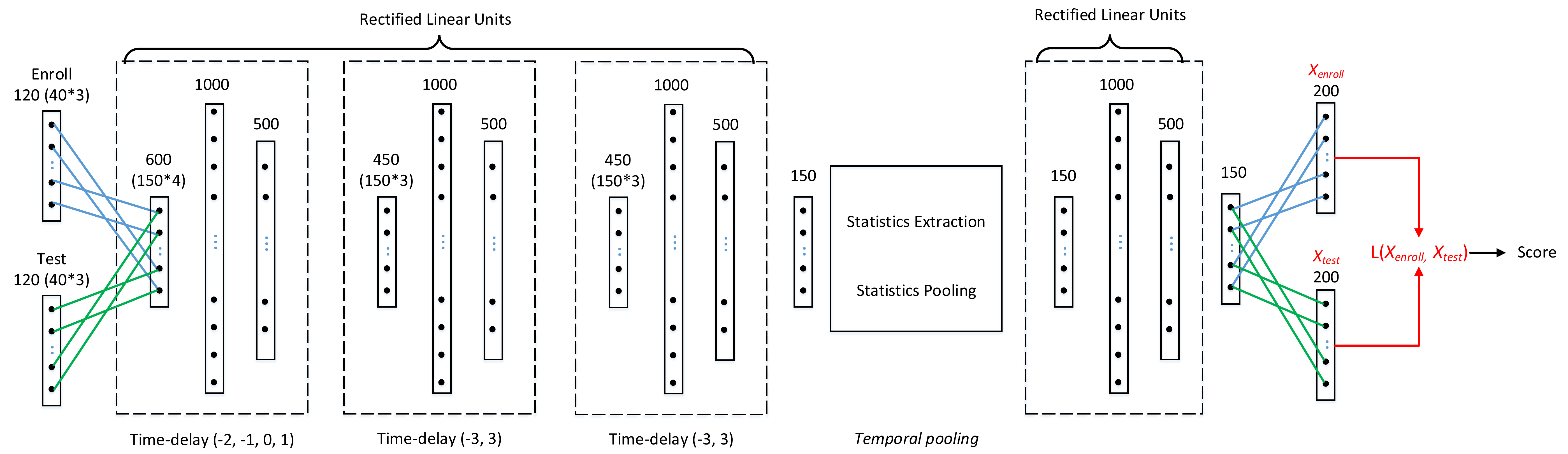}
\caption{The DNN structure of the end-to-end system~\cite{snyderdeep16}.}
\label{fig:end2end}
\end{figure*}

To perform SV, a simple back-end model is constructed, which consists of a simple average pooling
that averages the frame-level speaker features to utterance-level representations, denoted by `d-vectors',
and a scoring scheme based on the cosine distance between the d-vectors of the enrollment and test
utterances.

\subsection{End-to-end model}

The end-to-end DNN model proposed Snyder et al.~\cite{snyderdeep16} is used in our study.
A particular reason we choose this model is that it has been tested on text-independent tasks.
The model structure is illustrated in Fig.~\ref{fig:end2end}.
The input is a pair of utterances (in the form of feature sequences) sampled from the training data,
where the two utterances may be from the same speaker or from different speakers, labelled by $0$ and $1$
respectively.

The DNN structure consists of an embedding component and a scoring component (back-end).
The embedding component converts an input utterance to a speaker embedding. The utterance is first
propagated through three time-delay network-in-network (NIN) layers~\cite{ghahremani2016acoustic}.
Each NIN component is composed of a stack of three rectified linear units connected by affine transformations,
and maps the input that is $150$-dimensional to a $1,000$-dimensional space, then projects to the output
to a $500$-dimensional space.
The output of the third NIN layer is aggregated by a temporal pooling layer, by which the statistics of the input
utterance are derived.
Finally, the statistics are propagated to another NIN layer and a linear affine layer, producing the speaker embedding.
Note that in this work, we only use the mean vector as the statistics as it performed the best in our experiments.

The back-end scoring component estimates the probability that the two input utterances, represented by their embeddings,
belong to the same speaker. It is essentially a bi-linear projection followed by a logistic sigmoid function,
formulated as follows:

\[
   P_r(x,y)=\frac{1}{1 + e^{-L(x,y)}}
\]

\noindent where
\begin{equation}
        \label{equ:score}
    L(x,y) = x^Ty - x^TSx - y^TSy + b.
\end{equation}

By this DNN structure, the objective function of the training is simply the cross entropy between the
prediction of the network and the ground-truth of the training samples (pairs of utterances), formulated as:

\begin{equation}
   E = - \sum\limits_{(x,y)\in P_{same}}ln(P_r(x,y)) -  K \sum\limits_{(x,y)\in P_{diff}}ln(1 - P_r(x,y))
\end{equation}

\noindent where P$_{same}$ and P$_{diff}$ represent the set of same-speaker pairs and different-speaker pairs, respectively.
Since there are much more pairs in P$_{diff}$ than P$_{same}$, a constant hyper-parameter $K$ is introduced
to balance the contribution of each set.

\subsection{Comparison of feature learning and end-to-end}

The two deep SV approaches are fundamentally different from multiple aspects. A thorough comparison helps understand
their respective advantages.

\begin{itemize}

\item Difference in model structure. The end-to-end model involves both speaker embedding (front-end) and scoring (back-end), and the two components are trained jointly as an integrated network. The feature learning model, in contrast, involves only the front-end.

\item Difference in training objectives. The training objective of the end-to-end model is to directly determine if a pair of utterances
are from the same speaker or different speakers. The feature learning model, instead, aims to discriminate the speakers in the training set.
Obviously, the end-to-end objective is more consistent with the SV task.

\item Difference in training scheme. The end-to-end model is trained in a pair-wised style, which heavily relies on the quality and quantity of the sampled pairs. The feature learning model is trained in a one-hot style, for which a single training example triggers much stronger error signal through the softmax function. This suggests that
    training the feature learning model could be easier than training the end-to-end model, and requires less data and computation.

\item Difference in generalizability. The end-to-end approach is purely task-oriented, and the resultant system can perform SV only; the feature learning approach, instead, learns intermediate representations that can be used in a broad range of applications, such as
    speech signal factorization~\cite{wang2017deep}, speaker-dependent text-to-speech synthesis.

\end{itemize}

As a summary, the end-to-end model is theoretically optimal for SV, but the training could be difficult. The feature learning approach is opposite. Which approach is better in practical usage is therefore an open question.

\section{Experiments}
\label{sec:exp}

In this section, we first present the database and settings of different systems, then report the performance results.
Experiments were also conducted to analyze the factors that caused the different performance with the two deep SV systems.
All the experiments were conducted with the Kaldi toolkit~\cite{povey2011kaldi}.

\subsection{Database}
\label{sec:data}

Our experiments were conducted with the \emph{Fisher} database.
The training set and the evaluation set are presented as follows.

\begin{itemize}
    \item \textbf{Training set}: It consists of $2,500$ male and $2,500$ female speakers, with $95,167$ utterances randomly selected from the \emph{Fisher} database,
    and each speaker has about $20$ utterances and totally $120$ seconds in length.
    This dataset was used for training the i-vector system, LDA model, PLDA model, and the DNNs of two deep SV systems.
    \item \textbf{Evaluation set}: It consists of $500$ male and $500$ female speakers randomly selected from
    the \emph{Fisher} database. There is no overlap between the speakers of the training set and the evaluation set.

\end{itemize}

We set two test conditions: a short-enrollment condition (C(4-4)) and a
long-enrollment condition (C(40-4)), for which the duration of the enrollment utterances is $4$ seconds and
$40$ seconds, respectively. More details of the two test conditions are presented in Table~\ref{tab:data}.
The trials in the test are either female-female or male-male, and the results are reported on
the pool of all the trials.

\begin{table}[htp]
    \begin{center}
        \caption{Data profile of the test conditions.}
        \label{tab:data}
          \begin{tabular}{|l|c|c|}
           \hline
               Test condition   &   C(4-4)  &    C(40-4)    \\
           \hline
           \hline
               No. of Enrollment Utt.   &   82k     &    10k       \\
               No. of Test Utt.    &   82k     &    73k       \\
           \hline
               Avg. duration of Enrollment Utt.  &   4s      &    40s       \\
               Avg. duration of Test Utt.   &   4s      &    4s        \\
           \hline
                No. of Target Trials       &   3.5k    &    73k       \\
                No. of Non-target Trials    &   82M     &    36M       \\
           \hline
          \end{tabular}
    \end{center}
\end{table}

\subsection{Model settings}
\label{sec:model}

\subsubsection{I-vector system}
The i-vector system was built as a baseline for comparison. The raw feature involved $19$-dimensional MFCCs plus the log energy.
This raw feature was augmented by its first- and second-order derivatives, resulting in a
60-dimensional feature vector. This feature was used by the i-vector model.
The UBM was composed of $2,048$ Gaussian components, and the dimensionality of the i-vector space was $400$.
The dimensionality of the LDA projection space was set to $150$.
Prior to the PLDA scoring, the i-vectors were centered and length normalized. The entire system was trained using the Kaldi SRE08 recipe.

\subsubsection{Deep feature learning system}
The deep feature learning system was constructed based on the DNN structure shown in Fig.~\ref{fig:ctdnn}.
The input feature was 40-dimensional Fbanks, with a symmetric $4$-frame window to splice the neighboring frames,
resulting in $9$ frames in total.
With two time-delay hidden layers, the size of the effective context window is $20$ frames.
The number of output units was $5,000$, corresponding to the number of speakers in the training data.
The speaker features were extracted from the last hidden layer (the feature layer in Fig.~\ref{fig:ctdnn}),
and the utterance-level d-vectors were derived by averaging the frame-level features.
The three scoring metrics used by the i-vector system were also used by the d-vector system,
including cosine distance, LDA and PLDA.

\subsubsection{End-to-end system}
The end-to-end system was constructed based on the DNN architecture shown in Fig.~\ref{fig:end2end}.
The input feature was 40-dimensional Fbanks, with a symmetric $1$-frame window to splice the neighboring frames,
resulting in $3$ frames in total.
With three time-delay hidden layers, the size of the effective context window is $17$ frames.
The training samples were organized as pairs of feature chunks, which may be either same-speaker or different-speaker.
Each mini-batch involves $N$ same-speaker pairs and $N(N-1)$ different-speaker pairs.
Limited by the GPU memory, $N$ was set to $64$ in our experiment.
The number of frames in a feature chunk was a random variable sampled from a log-uniform distribution,
ranging from $50$ to $300$.
The feature dimensions before and after the temporal pooling are both $150$ (note that
the statistics produced by the temporal pooling is just the mean vector),
and the dimensionality of the speaker embedding was $200$.
In evaluation, the enrollment and test utterances were fed to the neural model simultaneously,
and then the decision scores were obtained from the output of the network.

We used the recipe published by the author of~\cite{snyderdeep16}, and tried our best to optimize the system
by tuning the configurations, including the trunk size and the batch size. The settings mentioned above are the optimal
values we found in the system tuning.

\subsection{Experimental results}
\label{sec:basic}

\begin{table}[htb]
    \begin{center}
      \caption{EER(\%) results of the three SV systems.}
      \label{tab:baseline}
      \footnotesize
          \begin{tabular}{|l|l|c|c|}
            \hline
            \multicolumn{2}{|c|}{}                &\multicolumn{2}{c|}{EER\%}\\
            \hline
               Systems              &   Scoring   &   C(4-4)   &    C(40-4)    \\
           \hline
           \hline
               i-vector             &    Cosine   &   16.96   &    4.81      \\
                                    &    LDA      &   10.95   &    3.30      \\
                                    &    PLDA     &   8.84    &    3.39      \\
            \hline
               Deep feature         &    Cosine   &   10.31   &    4.01      \\
                                    &    LDA      &\textbf{7.86} & \textbf{2.39}       \\
                                    &    PLDA     &   13.01   &    5.24      \\
            \hline
               End-to-end           &    -        &   9.85    &    4.59      \\
           \hline
          \end{tabular}
      \end{center}
\end{table}

The results of the three SV systems in terms of equal error rate (EER\%) are reported in Table~\ref{tab:baseline}.
Firstly we compare the three SV systems with their own best configurations, say, i-vector + PLDA, deep feature + LDA.
It can be seen that the deep feature system performs the best, and the end-to-end system is inferior compared to
the other two systems.
The relative better performance of the feature learning system compared to the i-vector system has been reported
in our previous paper~\cite{li2017deep}. The inferior performance of the end-to-end system compared to the i-vector
system with limited training data is also consistent with the results reported by Snyder et al.~\cite{snyderdeep16}, where
they found that the end-to-end model did not beat the best i-vector model (i-vector + PLDA) when the training set
contained $5,000$ speakers.

Another observation is that the LDA model plays an important role for the feature learning system. At the first glance,
this seems not reasonable, as the feature has been speaker discriminative, and the discriminative DNN training
should have superseded LDA. However, more careful analysis shows that LDA normalizes the within-class variation, which is
important for SV but not the goal of the DNN training. From this perspective, LDA can be regarded as a better
back-end model that learns an SV-oriented decision strategy. The importance of
an SV back-end model for d-vector systems was also noticed by Heigold et al.~\cite{heigold2016end}, who reported that a score normalization
(t-norm) is important to improve the performance of a d-vector system, although their study was based on a text-dependent task.
T-norm plays a similar role as LDA in normalization of within-class variation.

The probabilistic version of LDA (PLDA), however, does not provide any contribution (actually it hurts the performance).
The failure of PLDA has been reported in our previous work~\cite{li2017deep}.
A possible reason is that the mean d-vector of each speaker does not follow a Gaussian prior, so cannot be well modeled by the PLDA model.

These results demonstrated that although the end-to-end model is highly SV-oriented, it is not easy to take
full advantage of this model due to the difficulties in model training. In our experiments, we indeed
found that the end-to-end training is rather difficult: it requires careful tuning otherwise the training may divergent,
and much attention has to be paid on the training pair preparation, e.g., the number of training pairs in each iteration and
the number of frames in each training pair. We also found a non-linear dynamics during end-to-end training,
i.e., the objective function is stuck at a value for quite a long time, and suddenly
drops dramatically. This from another perspective demonstrated the difficulty of the end-to-end training.

\section{Conclusions}
\label{sec:conl}

This paper studied two deep speaker verification models. One is the end-to-end neural model and
the other is the deep feature learning model.
Our experimental results showed that the two deep speaker models achieved comparable or even better performance than the i-vector/PLDA model.
When comparing with each other, we found that the feature learning model performs better than the end-to-end model, although the latter
is assumed to be more consistent with the SV task. From these experiments, it seems that
the end-to-end learning is not very suitable for SV, at least with data and computation resources similar to our experiment.
Lots of questions remain open, e.g., how the two approaches will perform with the training set growing?
how to use the respective advantages of the two approaches to construct a more powerful deep speaker model?
All need careful investigation.

\section*{Acknowledgment}

This work was supported by the National Natural Science Foundation of China under Grant No. 61371136 / 61633013
and the National Basic Research Program (973 Program) of China under Grant No. 2013CB329302.

\bibliographystyle{IEEEtran}
\bibliography{mybib}

\end{document}